\documentclass[preprint]{elsarticle}

\usepackage{amsmath, amssymb, amsthm, soul}
\usepackage{color}
\usepackage{latexsym}
\usepackage{indentfirst}
\usepackage{graphicx}
\usepackage{placeins}
\usepackage{booktabs}
\usepackage{algorithm}
\usepackage{algorithmic}
\usepackage{multirow}
\usepackage{subfig}

\biboptions{compress}

\theoremstyle{plain}

\theoremstyle{definition}

\theoremstyle{remark}

\DeclareMathOperator{\Tr}{Tr}

\DeclareMathOperator{\spanop}{span}

\newcommand{\ud}{\,\mathrm{d}}

\newcommand{\Or}{\mathcal{O}}

\newcommand{\bd}[1]{\boldsymbol{#1}}
\newcommand{\wt}[1]{\widetilde{#1}}

\newcommand{\mc}[1]{\mathcal{#1}}

\newcommand{\abs}[1]{\left\lvert#1\right\rvert}

\newcommand{\average}[1]{\left\langle#1\right\rangle}

\newcommand{\jump}[1]{\big[\hspace{-0.7mm} \big[ #1 \big]
  \hspace{-0.7mm} \big]}
\newcommand{\mean}[1] {\big\{ \hspace{-0.7mm} \big\{ #1 \big\}
  \hspace{-0.7mm} \big\}}

\newcommand{\KS}{\mathrm{KS}}

\newcommand{\xc}{\mathrm{xc}}

\newcommand{\eff}{\mathrm{eff}}
\newcommand{\DG}{\mathrm{DG}}

\newcommand{\barint}{\kern4pt \raise3.4pt\hbox{\vrule height.6pt
    width7pt} \kern-11pt \int}

\journal{Journal of Computational Physics}

\begin{document}

\begin{frontmatter}

\title{Adaptive local basis set for Kohn-Sham density
functional theory in a discontinuous Galerkin
framework II: \\
Force, vibration, and molecular dynamics calculations}

\author[lbl]{Gaigong Zhang}

\author[ucb,lbl]{Lin Lin}
\ead{linlin@math.berkeley.edu}

\author[lbl]{Wei Hu}
\ead{whu@lbl.gov}

\author[lbl]{Chao Yang}
\ead{cyang@lbl.gov}

\author[lnl]{John E. Pask}
\ead{pask1@llnl.gov}

\address[ucb]{Department of Mathematics, University of California,
Berkeley, Berkeley, CA 94720.}

\address[lbl]{Computational Research Division, Lawrence Berkeley National
Laboratory, Berkeley, CA 94720.}

\address[lnl]{Physics Division, Lawrence Livermore National
Laboratory, Livermore, CA 94550.}

\begin{abstract}
Recently, we have proposed the adaptive local basis set for
electronic structure calculations based on Kohn-Sham density
functional theory in a pseudopotential framework. The adaptive local
basis set is efficient and systematically improvable for total
energy calculations. In this paper, we present the calculation of
atomic forces, which can be used for a range of applications such as
geometry optimization and molecular dynamics simulation. We
demonstrate that, under mild assumptions, the computation of atomic
forces can scale nearly linearly with the number of atoms in the
system using the adaptive local basis set. We quantify the accuracy
of the Hellmann-Feynman forces for a range of physical systems,
benchmarked against converged planewave calculations, and find that
the adaptive local basis set is efficient for both force and energy
calculations, requiring at most a few tens of basis functions per
atom to attain accuracy required in practice. Since the adaptive
local basis set has implicit dependence on atomic positions, Pulay
forces are in general nonzero. However, we find that the Pulay force
is numerically small and systematically decreasing with increasing
basis completeness, so that the Hellmann-Feynman force is sufficient
for basis sizes of a few tens of basis functions per atom. We verify
the accuracy of the computed forces in static calculations of
quasi-1D and 3D disordered Si systems, vibration calculation of a
quasi-1D Si system, and molecular dynamics calculations of H$_2$ and
liquid Al-Si alloy systems, where we find excellent agreement with
independent benchmark results in literature.
\end{abstract}


\end{frontmatter}


\section{Introduction}

Kohn-Sham density functional theory (KSDFT)~\cite{HohenbergKohn:64,
KohnSham:65} is the most widely used electronic structure model for
molecules and condensed matter systems. Kohn-Sham density functional
theory gives rise to a nonlinear eigenvalue problem, which is
commonly solved using the self-consistent field (SCF) iteration
method~\cite{Martin:04}. At each SCF step, a linear eigenvalue
problem with a fixed Kohn-Sham Hamiltonian defined by a fixed
electron density $\rho$ is solved. The solution to this linear
eigenvalue problem is used to update the electron density and
Kohn-Sham Hamiltonian in the SCF iteration. This is the most
computationally expensive part of the SCF iteration.  Although the
asymptotic complexity of the computation with respect to the number
of atoms depends on the algorithm used to solve the algebraic
eigenvalue problem, the prefactor, which is related to the number of
basis functions per atom, is characterized by how the problem is
discretized. Methods such as the planewave
method~\cite{PayneTeterAllenEtAl1992}, finite difference
method~\cite{ChelikowskyTroullierSaad1994}, and finite element
method~\cite{TsuchidaTsukada1995,PaskKleinFongSterne:99,
PaskSterne2005a,ChenDaiGongEtAl2014,BaoHuLiu2013}
exhibit systematic convergence with respect to the
number of basis functions per atom, but can require a large number
of basis functions per atom, from hundreds to thousands or more. The
number of degrees of freedom can be reduced by incorporating atomic
orbital physics into the basis~\cite{AverillEllis:73,
DelleyEllis:82, Eschrig:88, KoepernikEschrig:99, Kenny:00,
Junquera:01, Ozaki:03, BlumGehrkeHankeEtAl2009}. Compared to methods
such as the planewave method, however, it is more difficult to
improve the quality of such atomic-orbital basis in a systematic
fashion. The improvement can rely heavily on the practitioner's
experience with the underlying chemical system.

In a recent publication~\cite{LinLuYingE2012}, we presented a new
basis to discretize the Kohn-Sham Hamiltonian, called the adaptive
local basis (ALB). The basic idea is to partition the global domain
into a number of subdomains (called elements), and solve the
Kohn-Sham problem locally around each element to generate the basis
functions in each element. The basis so constructed is discontinuous
across element boundaries. Therefore, we use the discontinuous
Galerkin (DG) method~\cite{CockburnKarniadakisShu:00} to construct a
finite dimensional Kohn-Sham Hamiltonian in the discontinuous
representation. The DG approach for solving the Kohn-Sham equations is
also explored recently with enriched polynomial basis
functions~\cite{LuCaiXinEtAl2013}. Recently, the adaptive local basis functions has
been implemented in the DGDFT software
package~\cite{HuLinYang2015a}, which achieves massive
parallelization over $128,000$ cores with more than $80\%$ of
parallel efficiency for a two dimensional phospherene system
containing $14,000$ atoms. The solution produced by DGDFT is also
fully consistent with the solution of standard Kohn-Sham equations
in the limit of a complete basis set, and the error can be measured
by a posteriori error estimators~\cite{KayeLinYang2015}. We remark
that the idea of generating localized basis functions on the fly has
also been explored in other electronic structure software packages
such as ONETEP~\cite{SkylarisHaynesMostofiEtAl2005} and recently
BigDFT~\cite{MohrRatcliffBoulangerEtAl2014}, where localized basis
functions are continuous and are improved through an optimization
procedure. Filter diagonalization~\cite{RaysonBriddon2009} is
another approach for contracting basis functions and has been
applied to contract Gaussian type functions. The filter
diagonalization requires choosing trial functions, although the
choice of trial functions may not be straightforward for an initial
set of fine basis functions such as planewaves, finite elements or
wavelets.


In KSDFT, many quantities of interest can be obtained from the total
energy and atomic forces.  We have previously demonstrated the
effectiveness of the adaptive local basis~\cite{LinLuYingE2012} and
a variant, the element orbitals~\cite{LinYing2012}, for computing
the total energy of systems such as disordered bulk Na and Si,
graphene with defects, and edge reconstruction of large scale
armchair phospherene nanoribbon systems~\cite{HuLinYang2015}.  In
order to compute the atomic forces, which are given by the
derivatives of the total energy with respect to atomic positions,
the Hellmann-Feynman theorem~\cite{Hellmann:37,Feynman:39} is
typically employed, and the resulting force is called the
Hellmann-Feynman force. Since the adaptive local basis set depends
implicitly on the atomic positions, however, the atomic force and
Hellmann-Feynman force are in general not the same, and their
difference, the Pulay force~\cite{Pulay:69}, reflects the effect of
the atomic-position dependence of the basis. Although we have
recently demonstrated that the Pulay force can be eliminated
systematically through an additional optimization
procedure~\cite{LinLuYingE2012a}, the procedure can be costly,
especially in three-dimensional simulations. Therefore, it is of
interest to determine the extent to which the Pulay force is reduced
without such additional optimization as the size of the adaptive
local basis is increased, and to determine the size of basis
required for accurate quantum mechanical forces in practice.

Here, we describe the details  to compute Hellmann-Feynman forces in
the adaptive local basis, which can be evaluated with near linear
scaling cost with respect to the number of atoms, provided the
density matrix represented in the adaptive local basis is obtained.
We quantify the accuracy of the Hellmann-Feynman forces for a range
of systems compared to converged planewave calculations, and find
that the adaptive local basis set is accurate and efficient for both
energy and force calculations, achieving accuracies required in
practice with a few tens of basis functions per atom. We quantify
the Pulay force
for two test systems, including a quasi-1D disordered Si system and
a 3D disordered Si system.
We find that the magnitude of the Pulay force is readily reduced to $10^{-4}$ au
with a few tens of basis functions per atom.
To demonstrate that such accuracy is sufficient in practice, we
compute the vibrational frequencies for the quasi-1D disordered Si
system using a frozen phonon approach, and find that the vibrational
frequencies agree well with those obtained from converged planewave
calculations using ABINIT~\cite{Abinit1}. We further validate the
accuracy of the computed forces through molecular dynamics
simulations, and vibrational calculations of H$_{2}$ molecules
and pair-correlation functions of a liquid Al-Si alloy, and we find
excellent agreement with independent results in literature.

The remainder of this paper is organized as follows. Section~\ref{sec:dg} introduces the
discontinuous Galerkin framework for Kohn-Sham density functional theory
and the construction of the adaptive local basis functions.
Section~\ref{sec:force} discusses the computation of
the Hellmann-Feynman force.  We report numerical results
in Section~\ref{sec:numer}, followed by discussion and conclusions in
Section~\ref{sec:discussion}.


\section{Discontinuous Galerkin framework for Kohn-Sham density functional theory}
\label{sec:dg}


\subsection{Kohn-Sham density functional theory}

%
%

We consider a system consisting of $N_A$ nuclei and $N$ electrons. In
the Born-Oppenheimer approximation, for each set of nuclear
positions $\{R_{I}\}_{I=1}^{N_A}$, the electrons are relaxed to their ground
state. The ground state total energy is denoted by
$\mc{E}_{\text{tot}}(\{R_{I}\}_{I=1}^{N_A})$, and can be computed
in Kohn-Sham density functional
theory~\cite{HohenbergKohn:64,KohnSham:65} according to
\begin{equation}
  \mc{E}_{\text{tot}}(\{R_{I}\}_{I=1}^{N_A}) =\min_{\{\psi_i\}_{i=1}^{N}}
  \mc{E}_{\KS}(\{\psi_i\}_{i=1}^{N};\{R_{I}\}_{I=1}^{N_A}).
  \label{eqn:KSmin}
\end{equation}
For simplicity, we assume all quantities are
real, and neglect spin degeneracy, as well as temperature effects leading to
fractional occupation. We also omit the range of indices $I,i$
unless otherwise specified. $\mc{E}_{\KS}$ is the Kohn-Sham energy
functional and is given by
\begin{equation}\label{eqn:KSfunc}
  \begin{split}
     &\mc{E}_{\KS}(\{\psi_i\};\{R_{I}\}) \\
    =& \frac{1}{2} \sum_{i=1}^{N} \int \abs{\nabla
    \psi_i(x)}^2 \ud x + \int V_{\text{loc}}(x;\{R_{I}\}) \rho(x) \ud
    x\\
    &+ \sum_{i=1}^{N} \int \psi_{i}(x) V_{\text{nl}}(x,y;\{R_{I}\})
    \psi_i(y) \ud x \ud y
    +\frac{1}{2} \iint \frac{\rho(x) \rho(y)}{\abs{x-y}} \ud x \ud y \\
    &+ E_{\xc}[\rho] + \frac{1}{2}\sum_{I\ne J}
    \frac{Z_{I}Z_{J}}{\abs{R_{I}-R_{J}}}.
  \end{split}
\end{equation}
We include the $\{R_I\}$ dependence explicitly to facilitate the
derivation of atomic forces.
Here,
\begin{equation}
    \label{eqn:KSden}
    \rho(x) = \sum_{i=1}^{N} \abs{\psi_i(x)}^2
\end{equation}
is the electron density.
The eigenfunctions (also called Kohn-Sham orbitals)
$\{\psi_i\}$ satisfy the orthonormality constraints
\begin{equation}
  \int \psi_i(x) \psi_j(x) \ud x = \delta_{ij}.
\end{equation}
In \eqref{eqn:KSfunc}, we use a norm-conserving
pseudopotential~\cite{Martin:04}.
The term
\begin{equation}
  V_{\text{loc}}(x;\{R_{I}\}) = \sum_{I=1}^{N_A}
  V_{\text{loc},I}(x-R_{I})
  \label{eqn:localpot}
\end{equation}
is the local part of the pseudopotential. Each term
$V_{\text{loc},I}(x-R_{I})$ is centered on the $I$-th atom, and decays asymptotically
as $-\frac{Z_{I}}{|x-R_{I}|}$ for large $|x-R_{I}|$,
where $Z_{I}$ is the charge of the $I$-th nucleus.
The nonlocal part of the pseudopotential takes the Kleinman-Bylander
form~\cite{KleinmanBylander:82}
\begin{equation}
  V_{\text{nl}}(x,y;\{R_{I}\}) =
  \sum_{I=1}^{N_A}\sum_{\ell=1}^{L_{I}} \gamma_{I,\ell}
  b_{I,\ell}(x-R_{I}) b_{I,\ell}(y-R_{I}).
  \label{eqn:nonlocalpot}
\end{equation}
For each atom $I$, there are $L_{I}$ functions $\{b_{I,\ell}\}$ called
projectors of the nonlocal pseudopotential.  Each $b_{I,\ell}$ is
centered at $R_{I}$ and is supported locally in real space
around $R_{I}$. $\gamma_{I,\ell}$ is a real scalar.
$E_{\xc}$ is the exchange-correlation energy. Here, we assume local or
semi-local exchange-correlation functionals are used. The last term in
Eq.~\eqref{eqn:KSfunc} is the ion-ion Coulomb interaction energy.
We note that for extended systems, modeled as infinite periodic structures,
both the local-pseudopotential and ion-ion terms require
special treatment in order to avoid divergences
due to the long-range $1/r$ nature of the Coulomb interaction.
We provide corresponding expressions for this case in the appendix A.

When the atomic positions $\{R_{I}\}$ are fixed, we may simplify the notation and
drop the $\{R_{I}\}$-dependence in $V_{\text{loc}}$ and $V_{\text{nl}}$.  The
Kohn-Sham equation is the Euler-Lagrange equation associated with
\eqref{eqn:KSfunc}:
\begin{equation}\label{eqn:KSeqn}
  H_{\eff}[\rho] \psi_i = \left( - \frac{1}{2} \Delta + V_{\eff}[\rho]
  + V_{\text{nl}}\right)
  \psi_i = E_i \psi_i.
\end{equation}
Here the effective single-particle potential $V_{\eff}$ is defined as
\begin{equation}\label{eqn:Veff}
  V_{\eff}[\rho](x) = V_{\text{loc}}(x) + V_{H}(x)
  + V_{\xc}[\rho](x),
\end{equation}
in which the Coulomb potential is given by
\begin{equation}
  V_{H}(x) = \int \frac{\rho(y)}{\abs{x - y}}\ud y.
  \label{eqn:Vcoulomb}
\end{equation}
$V_{\xc}[\rho](x)=\frac{\delta E_{\xc}}{\delta \rho}(x)$ is the
exchange-correlation potential.
Note that Eq.~\eqref{eqn:KSeqn} is a nonlinear eigenvalue problem, as
$V_{\eff}$ depends on $\rho$, which is in turn determined by
$\{\psi_i\}$.  The electron density is self-consistent if
both~\eqref{eqn:KSden} and ~\eqref{eqn:KSeqn} are satisfied.
After obtaining the self-consistent
electron density, the total
energy of the system can be expressed using the eigenvalues $\{E_{i}\}$ and
density $\rho$ as~\cite{Martin:04}
\begin{equation}
  \begin{split}
    \mc{E}_{\text{tot}} =&
    \sum_{i=1}^{N} E_{i} - \frac12 \iint \frac{\rho(x) \rho(y)}{\abs{x-y}} \ud x \ud y
    + E_{\xc}[\rho] - \int V_{\xc}[\rho](x) \rho(x) \ud x \\
    & + \frac{1}{2}\sum_{I\ne J}
    \frac{Z_{I}Z_{J}}{\abs{R_{I}-R_{J}}}.
  \end{split}
  \label{eqn:Etot}
\end{equation}

In the self-consistent field iteration for solving the Kohn-Sham equations,
the total computational time is usually dominated by the following
step: Given an input electron density and associated effective potential $V_{\eff}(x)$,
we find the output electron density $\wt{\rho}(x)$ from
\begin{equation}
  \wt{\rho}(x) = \sum_{i=1}^N \abs{\psi_i(x)}^2,
\end{equation}
where $\{\psi_i\}$ are the lowest $N$ eigenfunctions of $H_{\eff}$
in Eq.~\eqref{eqn:KSeqn}.
The $\{\psi_i\}$ then minimize the quadratic energy functional
\begin{equation}\label{eqn:linearvar}
  \begin{split}
  \mc{E}_{\eff}(\{\psi_i\}) = &\frac{1}{2} \sum_{i=1}^N \int
  \abs{\nabla \psi_i(x)}^2 \ud x + \sum_{i=1}^{N} \int V_{\eff}(x)
  \abs{\psi_{i}(x)}^2 \ud x \\
  &+ \sum_{i=1}^{N} \int \psi_{i}(x) V_{\text{nl}}(x,y) \psi_i(y)
  \ud x \ud y,
  \end{split}
\end{equation}
with $\{\psi_i\}$ being orthonormal. Note that the ion-ion interaction is
a constant depending only on $\{R_{I}\}$, and is dropped in
Eq.~\eqref{eqn:linearvar}.

\subsection{Adaptive local basis and discontinuous Galerkin framework}

In~\cite{LinLuYingE2012}, the adaptive local basis functions in a
discontinuous Galerkin (DG) framework have been proposed to reduce the
computational time for solving the equations of KSDFT. The DG method relaxes the
continuity constraint on basis functions, and provides flexibility in
choosing the basis set for efficient discretization.  Among the
different formalisms in the DG framework, we use the interior penalty
method~\cite{BabuskaZlamal:73,Arnold:82}, which naturally generalizes
the variational principle~\eqref{eqn:linearvar}.

We denote by $\Omega$ the computational domain with periodic boundary
conditions, which corresponds to $\Gamma$ point sampling in the Brillouin
zone~\cite{Martin:04}. The domain $\Omega$ is also referred to as the global domain in the following.
More general Bloch boundary conditions may be accommodated as well.
Let $\mc{T}$ be a collection of quasi-uniform rectangular partitions of $\Omega$,
\begin{equation}
  \mc{T} = \{E_1, E_2, \cdots, E_M \},
\end{equation}
and $\mc{S}$ be the collection of surfaces that correspond to
$\mc{T}$.  Each $E_{k}$ is called an element of $\Omega$. For a
typical choice of partitions used in practice, the elements are chosen
to be of the same size.

We define the following inner products:
\begin{align}
  & \average{v, w}_E = \int_E v(x) w(x) \ud x, & \average{\bd{v}, \bd{w}}_S = \int_S \bd{v}(x)
  \cdot \bd{w}(x) \ud s(x), \\
  & \average{v, w}_{\mc{T}} = \sum_{i = 1}^M
  \average{v, w}_{E_i}, & \average{\bd{v}, \bd{w}}_{\mc{S}} = \sum_{S \in \mc{S}}
  \average{\bd{v}, \bd{w}}_S.
\end{align}
In the interior penalty method, the energy functional
corresponding to \eqref{eqn:linearvar} is given by
\begin{equation}\label{eqn:DGvar}
  \begin{split}
    E_{\DG}(\{\psi_i\}) = &\frac{1}{2} \sum_{i=1}^N \average{\nabla
    \psi_i , \nabla \psi_i}_{\mc{T}}
    + \average{ V_{\eff}, \rho }_{\mc{T}}
    + \sum_{I=1}^{N_A}\sum_{\ell=1}^{L_{I}} \gamma_{I,\ell} \sum_{i=1}^N
    \abs{\average{b_{I,\ell}(\cdot-R_{I}), \psi_i}_{\mc{T}}}^2 \\
    &- \sum_{i=1}^N
    \average{\mean{\nabla\psi_i}, \jump{\psi_i}}_{\mc{S}}
    + \alpha \sum_{i=1}^N \average{\jump{\psi_i},
    \jump{\psi_i}}_{\mc{S}}.
  \end{split}
\end{equation}
Here, $\mean{\cdot}$ and $\jump{\cdot}$ are the average and the jump
operators across surfaces, defined as follows. Because of the periodic
boundary condition, each surface $S\in \mc{S}$ is an interior surface in
the sense that $S$ is shared by elements $K_1$ and $K_2$.  Denote by
$n_1$ and $n_2$ the unit normal vectors on $S$ pointing exterior to
$K_1$ and $K_2$, respectively. With $u_i = u\vert_{\partial K_i}$, $i =
1, 2$, we set
\begin{equation}
  \jump{u} = u_1 n_1 + u_2 n_2 \quad \text{on } S \in \mc{S}.
\end{equation}
For vector-valued
function $q$, we define
\begin{equation}
  \mean{q} = \frac{1}{2} (q_1 + q_2) \quad \text{on } S \in \mc{S},
\end{equation}
where $q_i = q \vert_{\partial K_i}$.
The second to last term in Eq.~\eqref{eqn:DGvar} comes from
integration by parts of the Laplacian operator, which cures the
ill-defined operation of applying the Laplacian operator to
discontinuous functions in order to define the kinetic energy.
The last term in
Eq.~\eqref{eqn:DGvar} is a penalty term which penalizes the jumps of
functions across element surfaces to guarantee
stability~\cite{Arnold:02}, and the constant $\alpha$  is a positive
penalty parameter.  We have demonstrated that the adjustable penalty
parameter $\alpha$ is mainly used to ensure the stability of the numerical
scheme, and has relatively little effect of the accuracy of the scheme
when it takes a large range of values~\cite{LinLuYingE2012,HuLinYang2015a}.

Assume that we have chosen for each element $E_k$ a set of basis
functions $\{\varphi_{k,j}\}_{j=1}^{J_k}$, where $J_k$ is the number
of basis functions in $E_k$. We extend each $\varphi_{k,j}$ to the
whole computational domain $\Omega$ by setting it to zero on the
complement set of $E_k$. Define the function space $\mc{V}$ as
\begin{equation}
  \mc{V} = \spanop\{ \varphi_{k,j},\, k=1,\cdots,M;\, j = 1, \cdots, J_k \}.
\end{equation}
The local basis functions $\{\varphi_{k,j}\}_{j=1}^{J_k}$ which we use to
discretize the Kohn-Sham problem are constructed as
follows.  For each $E_{k}\in \mc{T}$, we introduce an associated {\em extended
element} $Q_{k} \supset E_{k}$, with $Q_{k} \backslash E_{k}$ a buffer
region surrounding $E_{k}$.  We define $V_{\eff}^{Q_{k}}=V_{\eff}\vert_{Q_{k}}$
to be the restriction of the effective potential at the current SCF
step to $Q_{k}$, and
$V_{\text{nl}}^{Q_{k}}=V_{\text{nl}}\vert_{Q_{k}}$ to be the restriction
of the nonlocal potential to $Q_{k}$. We solve the local eigenvalue
problem on each extended element
\begin{equation}
    \left(-\frac12 \Delta + V_{\eff}^{Q_{k}}
    +V_{\text{nl}}^{Q_{k}}\right) \wt{\varphi}_{k,j} =
    \lambda_{k,j} \wt{\varphi}_{k,j}.
    \label{eqn:localproblem}
\end{equation}
The lowest $J_{k}$ eigenvalues
$\{\lambda_{k,j}\}_{j=1}^{J_{k}}$ and corresponding orthonormal eigenfunctions
$\{\wt{\varphi}_{k,j}\}_{j=1}^{J_{k}}$ are computed.
We then restrict $\{\wt{\varphi}_{k, j}\}_{j=1}^{J_{k}}$ from $Q_{k}$ to
$E_{k}$. The truncated functions are not necessarily orthonormal.
Therefore, we apply a singular value decomposition (SVD)
to obtain $\{\varphi_{k,j}\}_{j=1}^{J_k}$. The
SVD procedure can ensure the orthonormality
of the basis functions inside each element,
as well as eliminating the linearly  dependent and nearly linearly
dependent functions in the basis set.  We then extend each
$\varphi_{k,j}$ to the global domain by setting it to zero
outside of $E_{k}$, so that it is in general
discontinuous across the boundary of $E_{k}$. As a result, the overlap
matrix corresponding to the adaptive local basis set is an identity
matrix.

There are a number of possible ways to set the boundary conditions for
the local problem~\eqref{eqn:localproblem}.
In practice, we use periodic boundary conditions for all eigenfunctions
$\{\wt{\varphi}_{k,j}\}_{j=1}^{J_{k}}$ in $Q_{k}$.
In some sense, the details of boundary condition do not
affect the accuracy of the adaptive local basis set much as the buffer
size increases.
This permits the use
of highly efficient Fourier based solution methods for the local problem.
The size of each extended
element should be chosen to balance between the effectiveness of the
basis functions and the computational cost for obtaining them.
For a typical choice used in practice, the elements are chosen to be of the same size, and
each element contains on average a few atoms.  The partition does not
need to be updated when the atomic configuration is changed, as in
the case of structure optimization and molecular dynamics.

After obtaining the basis functions, we minimize~\eqref{eqn:DGvar} for
$\{ \psi_i\} \subset \mc{V}$, i.e.,
\begin{equation}
  \psi_i(x) = \sum_{k=1}^{M} \sum_{j=1}^{J_k} c_{i; k, j}
  \varphi_{k, j}(x).
  \label{eqn:psiexpand}
\end{equation}
The output electron density is then computed as
\begin{equation}\label{eqn:newdensity}
  \wt{\rho}(x) = \sum_{i=1}^N \sum_{k=1}^{M} \abs{\sum_{j=1}^{J_k}
  c_{i; k, j} \varphi_{k, j}(x)}^2.
\end{equation}
Note that the computation of the electron density can be performed
locally in each element, since for each $x\in \Omega$ there is a unique
$k\equiv k(x)$ such that $\varphi_{k,j}(x)\ne 0$.  We refer readers
to~\cite{LinLuYingE2012} for details of solving the minimization
problem~\eqref{eqn:DGvar} as an eigenvalue problem in the DG
formulation.

\section{Calculation of atomic forces}\label{sec:force}

Once the SCF iteration reaches convergence to yield converged electron
density $\rho(x)$ and Kohn-Sham orbitals $\{\psi_{i}\}$, the force on the
$I$-th atom can be
computed as the negative derivative of the total energy with respect to
the atomic position $R_{I}$:
\begin{equation}
  F_{I} = -\frac{\partial \mc{E}_{\text{tot}}(\{R_{I}\})}{\partial
  R_{I}}.
  \label{eqn:forceDef}
\end{equation}
The required derivative can be computed directly, e.g., via finite differences.
However, even for first order accuracy,
the number of energy evaluations for a system containing $N_A$
atoms is $3N_A+1$, i.e., the Kohn-Sham equations must be solved
$3N_A+1$ times independently.  This approach becomes prohibitively expensive as
the system size increases.
The cost of the force calculation is
greatly reduced via the Hellmann-Feynman theorem,
which states that, at self-consistency,
the partial derivative $\frac{\partial}{\partial R_{I}}$ only needs to be applied to
terms in Eq.~\eqref{eqn:KSfunc} which depend \textit{explicitly} on the
atomic position $R_{I}$. The Hellmann-Feynman (HF) force is then given by
\begin{equation}
  \begin{split}
    F_{I}^{\text{HF}} = &-\int \frac{\partial
    V_{\text{loc}}}{\partial R_{I}}(x;\{R_{I}\}) \rho(x) \ud x
    - \sum_{i=1}^{N} \int \psi_{i}(x)
    \frac{\partial V_{\text{nl}}}{\partial R_{I}}(x,y;\{R_{I}\})
    \psi_{i}(y) \ud x \ud y\\
    &+ \sum_{J\ne I}\frac{Z_{I}Z_{J}}{\abs{R_{I}-R_{J}}^3}(R_{I}-R_{J}).
  \end{split}
  \label{eqn:forceHF1}
\end{equation}
Note that Eq.~\eqref{eqn:localpot} gives
\[
  \frac{\partial V_{\text{loc}}}{\partial R_{I}}(x;\{R_{I}\}) =
  \frac{\partial V_{\text{loc},I}}{\partial R_{I}}(x-R_{I}) =
  -\nabla_{x} V_{\text{loc},I}(x-R_{I}),
\]
and similarly Eq.~\eqref{eqn:nonlocalpot} gives
\[
  \begin{split}
    &\frac{\partial V_{\text{nl}}}{\partial R_{I}}(x,y;\{R_{I}\})\\
    = &\sum_{\ell=1}^{L_{I}} \gamma_{I,\ell} \left( \frac{\partial
    b_{I,\ell}}{\partial R_{I}}(x-R_{I}) b_{I,\ell}(y-R_{I}) +
    b_{I,\ell}(x-R_{I}) \frac{\partial b_{I,\ell}}{\partial
    R_{I}}(y-R_{I})\right)\\
    =& -\sum_{\ell=1}^{L_{I}}
    \gamma_{I,\ell}
    \left( \nabla_{x} b_{I,\ell}(x-R_{I}) b_{I,\ell}(y-R_{I}) +
    b_{I,\ell}(x-R_{I}) \nabla_{y} b_{I,\ell}(y-R_{I}) \right).
  \end{split}
\]
Then the Hellmann-Feynman force in Eq.~\eqref{eqn:forceHF1} can be written as
\begin{equation}
  \begin{split}
    F_{I}^{\text{HF}} = &\int \nabla_{x} V_{\text{loc},I}(x-R_{I})
    \rho(x) \ud x\\
    & + 2\sum_{i=1}^{N} \sum_{\ell=1}^{L_{I}} \gamma_{I,\ell}
    \left(\int \psi_{i}(x) \nabla_{x} b_{I,\ell}(x-R_{I}) \ud x \right)
    \left(\int \psi_{i}(y) b_{I,\ell}(y-R_{I}) \ud y \right)\\
    & + \sum_{J\ne I}\frac{Z_{I}Z_{J}}{\abs{R_{I}-R_{J}}^3}(R_{I}-R_{J}).
  \end{split}
  \label{eqn:forceHF2}
\end{equation}

From the computational cost point of view, if we denote by $N_{g}$ the
number of grid points to discretize quantities such as $\rho(x)$ in the
global domain, then the cost of computing each integral in the form
$\int \nabla_{x} V_{\text{loc},I}(x-R_{I}) \rho(x) \ud x$ is
$\Or(N_{g})$, since $V_{\text{loc},I}(x-R_{I})$ is a
delocalized quantity in the global domain.  On the other hand, each
nonlocal projector $b_{I,\ell}(x-R_{I})$ is localized around
$R_{I}$, and the cost of evaluating the integral $\left(\int
\psi_{i}(x) \nabla_{x} b_{I,\ell}(x-R_{I}) \ud x \right)$ or $\left(\int
\psi_{i}(y) b_{I,\ell}(y-R_{I}) \ud y \right)$ is a constant
$N_{l}$ independent of the global number of grid points $N_{g}$. The
computation of the last term  $\sum_{J\ne
I}\frac{Z_{I}Z_{J}}{\abs{R_{I}-R_{J}}^3}(R_{I}-R_{J})$ involves only
scalar operations, and its cost is usually negligibly small in
electronic structure calculations.
$N_{g}$ and $N_{A}$ are proportional to the number of electrons $N$.
Hence, neglecting constant terms independent of $N$,
we have that the computational cost of
the Hellmann-Feynman force on each atom is $\Or(N_{g} + N L_{I} N_{l})
\sim \Or(N)$, so that the cost for all atoms is
$\Or(N^{2})$.

We now demonstrate that with auxiliary quantities, the
cost of computing the Hellmann-Feynman forces on all atoms can be reduced to $\Or(N)$
without loss of accuracy in the DG formulation.  The strategy is
different for the local pseudopotential term and nonlocal
pseudopotential term, respectively.

For the local term, let us rewrite
$V_{\text{loc},I}(x-R_{I})$ as
\begin{equation}
  V_{\text{loc},I}(x-R_{I}) = -\int
  \frac{\rho_{\text{loc},I}(y-R_{I})}{\abs{x-y}} \ud y.
  \label{eqn:pseudocharge}
\end{equation}
The term $\rho_{\text{loc},I}$ is called the ionic pseudocharge density, or smeared
ionic density corresponding to the local
pseudopotential~\cite{SolerArtachoGaleEtAl2002,PaskSterne2005}, and the minus sign
in Eq.~\eqref{eqn:pseudocharge}
reflects the opposite sign of electronic and ionic charge.
While $V_{\text{loc},I}$ is delocalized in the global domain due to the long-range
Coulomb interaction, the corresponding pseudocharge $\rho_{\text{loc},I}$ is
localized around the nuclear position $R_{I}$, similar to the projector $b_{I,\ell}$ of
the nonlocal pseudopotential. Then,
\begin{equation}
  \begin{split}
    \int \nabla_{x} V_{\text{loc},I}(x-R_{I}) \rho(x) \ud x
    = &\int \nabla_{x} \left(-\int
    \frac{\rho_{\text{loc},I}(y-R_{I})}{\abs{x-y}} \ud y\right) \rho(x) \ud x \\
    = & \int \rho_{\text{loc},I}(y-R_{I}) \nabla_{y} \left(
    \frac{1}{\abs{x-y}}\right) \rho(x) \ud x \ud
    y \\
    = & \int \rho_{\text{loc},I}(y-R_{I}) \nabla_{y} \left(
    \int \frac{\rho(x)}{\abs{x-y}} \ud x \right)  \ud y  \\
    = & \int \rho_{\text{loc},I}(y-R_{I}) \nabla_{y} V_{H}(y) \ud y.
  \end{split}
  \label{eqn:localforce}
\end{equation}
Here, $V_{H}$ is the Coulomb potential defined in
Eq.~\eqref{eqn:Vcoulomb}.  Note that $V_{H}$ and its gradient only needs
to be evaluated once for all atoms.  Due to the locality of the ionic
pseudocharge, the cost of numerical integration using the last equality in
Eq.~\eqref{eqn:localforce} is independent of the global number of grid
points $N_{g}$ and number of electrons $N$.  So the computation of
the local pseudopotential part of the Hellmann-Feynman force for all
atoms scales as $\Or(N)$.

For the nonlocal part of the pseudopotential, the cost arises from the fact that
all $N$ Kohn-Sham orbitals need to participate in the evaluation of the
force for each atom $I$.  The cost can be reduced by an alternative
formulation using the density matrix, defined as
\begin{equation}
  P(x,y) = \sum_{i=1}^{N} \psi_{i}(x) \psi_{i}(y).
\end{equation}
Recalling the expansion of the Kohn-Sham orbitals in
Eq.~\eqref{eqn:psiexpand} in terms of the adaptive local basis
functions, the density matrix can also be expanded as
\[
P(x,y) = \sum_{i=1}^{N} \sum_{k=1}^{M} \sum_{j=1}^{J_{k}}
\sum_{k'=1}^{M} \sum_{j'=1}^{J_{k'}} \varphi_{k,j}(x)
\varphi_{k',j'}(y) c_{i; k,j} c_{i; k', j'}.
\]
Omitting the range of summation for $k,j,k',j'$,
the nonlocal part of the force becomes
\begin{equation}
  \begin{split}
    &2\sum_{i=1}^{N} \sum_{\ell=1}^{L_{I}}
    \left(\int \psi_{i}(x) \nabla_{x} b_{I,\ell}(x-R_{I}) \ud x \right)
    \left(\int \psi_{i}(y) b_{I,\ell}(y-R_{I}) \ud y \right)\\
   =&2 \sum_{\ell=1}^{L_{I}} \int \nabla_{x} b_{I,\ell}(x-R_{I}) P(x,y)
    b_{I,\ell}(y-R_{I}) \ud x \ud y\\
    =&2 \sum_{\ell=1}^{L_{I}} \sum_{k,j,k',j'}
    \left(V_{\text{nl},I,\ell}\right)_{k,j;k',j'} P_{k,j;k',j'}
    \equiv 2 \sum_{\ell=1}^{L_{I}} \Tr[V_{\text{nl},I,\ell} P].
  \end{split}
  \label{eqn:nonlocaldm}
\end{equation}
Here,
\[
\left(V_{\text{nl},I,\ell}\right)_{k,j;k',j'} = \left(\int \varphi_{k,j}(x) \nabla_{x}
b_{I,\ell}(x-R_{I}) \ud x \right) \left(\int \varphi_{k',j'}(y)
b_{I,\ell}(y-R_{I}) \ud y\right)
\]
is the nonlocal pseudopotential matrix element for the $\ell$-th nonlocal
pseudopotential projector of the $I$-th atom, represented in the adaptive local basis
set, and
\[
P_{k,j;k',j'} = \sum_{i=1}^{N} c_{i;k,j} c_{i;k',j'}
\]
is the corresponding density matrix element represented in the adaptive local basis
set.
Since the basis functions and projectors are localized, the cost of evaluating each
matrix element of $V_{\text{nl},I,\ell}$ is independent of $N$, and the number of
nonzeros in $V_{\text{nl},I,\ell}$ is independent of $N$ as well. Thus the cost
of evaluating the trace operation in Eq.~\eqref{eqn:nonlocaldm} for each atom $I$
is independent of $N$.
Therefore,
if the density matrix represented in the adaptive local basis set is
computed, the computational cost of the nonlocal pseudopotential component of the force
for all atoms scales as $\Or(N)$ as well.  We remark that diagonalization-free
numerical methods for computing the density matrix represented in the
adaptive local basis set are available. They include linear scaling
methods~\cite{Goedecker1999,BowlerMiyazaki2012} for
insulating systems, and the recently developed pole expansion and
selected inversion (PEXSI)
method~\cite{LinGarciaHuhsEtAl2014,LinChenYangEtAl2013,LinLuYingEtAl2009} for both
insulating and
metallic systems.
The DG Hamiltonian matrix corresponding to the adaptive local basis set
has a block stencil like sparsity structure and the overlap matrix is an
identity matrix, and is therefore well suited for such diagonalization
free methods.

In sum, the Hellmann-Feynman force can be compactly written and computed
in the DG formulation as
\begin{equation}
  \begin{split}
    F_{I}^{\text{HF}} = &\int \rho_{\text{loc},I}(x-R_{I}) \nabla_{x} V_{H}(x) \ud x
     + 2 \sum_{\ell=1}^{L_{I}}\Tr[V_{\text{nl},I,\ell} P]  \\
     &+ \sum_{J\ne
    I}\frac{Z_{I}Z_{J}}{\abs{R_{I}-R_{J}}^3}(R_{I}-R_{J}).
  \end{split}
  \label{eqn:hfforce}
\end{equation}
The computational cost to obtain the forces on all atoms is $\Or(N)$,
provided that the density matrix represented in the
adaptive local basis set is computed, neglecting the small cost of the ion-ion term.

In general, the force $F_{I}$ in Eq.~\eqref{eqn:forceDef} and
Hellmann-Feynman force $F_{I}^{\text{HF}}$ in Eq.~\eqref{eqn:hfforce} are not the same
when the basis used in discretization depends on $R_I$.
But they can be the same if one of the following three conditions is satisfied
\begin{enumerate}
  \item The basis set is complete.
  \item The basis set is not complete, but is independent of the atomic
    positions.  This is the case, e.g., for planewave basis functions.
  \item The basis set is not complete, but the basis set can be embedded
    in a larger subspace which is independent of atomic positions, and
    the basis set achieves the minimal energy among all choices of basis
    sets within the larger subspace.  This is the case, e.g., for the
    recently proposed optimized
    local basis (OLB) functions~\cite{LinLuYingE2012a}, which is a variant of
    the ALB in the DG framework.
\end{enumerate}

If none of the conditions above is satisfied, as e.g., for Gaussian
type orbitals (GTO) or atomic orbitals (AO), an additional term
called the Pulay force~\cite{Pulay:69} (denoted by $F_{I}^{\text{P}}$)
is needed so that
\begin{equation}
  F_{I} = F_{I}^{\text{HF}}+F_{I}^{\text{P}}.
  \label{eqn:Pulay}
\end{equation}
Since the Pulay force arises from the
$\{R_{I}\}$-dependence of the basis, it is present
for the adaptive local basis also.
However, due to the local optimality of the construction in the vicinity of each element,
the Pulay force is small and readily reduced to chemical accuracy,
as we show below.


\section{Results and discussion}\label{sec:numer}




In order to ascertain the accuracy and convergence of the obtained
Hellmann-Feynman forces in the adaptive local basis, we consider a
range of test cases, including quasi-1D and 3D, metallic and
insulating. We first consider static configurations, where we
compare directly to converged planewave calculations to determine
the absolute accuracy and convergence of the computed forces and
size of the Pulay component as the number of ALBs is increased. We
then employ the forces so obtained to compute vibrational
frequencies using a frozen phonon approach and pair correlation
functions from \textit{ab initio} molecular dynamics simulations,
comparing to independent calculations and previous work.

The ALB and DG based calculation is carried out using our new
massively parallel code called DGDFT~\cite{HuLinYang2015a}. Along
each dimension ($x,y,z$), the size of the extended element is fixed
to be $3$ times the size of the element, unless there is only $1$
element along this dimension. Since periodic boundary conditions are
assumed for the extended elements, we use the planewave basis set to
expand the ALBs in the extended element. It follows from the
standard planewave convention that the number of grid points in the
extended element is determined by the kinetic energy cutoff denoted
by $E_{\text{cut}}$. The relationship between $E_{\text{cut}}$ and
the number of uniform grid points along the $i$th direction ($N_i$),
where $i\in\{x,y,z\}$,
 can be written as
\begin{equation}
N_{i} = \frac{\sqrt{2E_{\text{cut}}}L_{i}}{\pi}.
  \label{eqn:Ecut}
\end{equation}
where $L_{i}$ is the dimension of the extended element along the $i$th
direction.  Since the wavefunctions are generally smoother than the
electron density and potential, a second set of uniform grids is employed to
accurately represent the density and potential with a higher
kinetic energy cutoff $E_{\text{cut}}^{\text{den}}$. In all of our calculations,
we set $E_{\text{cut}}^{\text{den}} = 4 E_{\text{cut}}$. In addition
to the wavefunction and density grids, we generate another grid which we
call the Legendre-Gauss-Lobatto (LGL) grid on each element
to perform accurate numerical quadrature.
To be consistent in notation, the number of LGL grid
points along each dimension is also defined in terms of a kinetic energy
cutoff denoted by $E_{\text{cut}}^{\text{LGL}}$.  We note that
$E_{\text{cut}}^{\text{LGL}}$ does not carry any physical meaning
for LGL grids, but is merely used as a convenient notation for fixing
the number of grid points from an equation analogous to Eq.~\eqref{eqn:Ecut}.
More details of the implementation in DGDFT can be found
in~\cite{LinLuYingE2012,HuLinYang2015a}.

For all calculations here, we use the
local density approximation
(LDA)~\cite{CeperleyAlder1980,PerdewZunger1981} for exchange and correlation,
and Hartwigsen-Goedecker-Hutter
(HGH)~\cite{HartwigsenGoedeckerHutter1998} pseudopotentials to model the ions.

We use the ABINIT planewave DFT code~\cite{Abinit1} as reference, to assess the
accuracy of our calculations.
The same exchange-correlation functionals and pseudopotentials are employed
in ABINIT and DGDFT, so that results can be compared directly, and errors assessed rigorously.

All calculations were carried out on the Hopper and Edison
systems at the National Energy Research Scientific Computing
Center (NERSC). There are 24 processors on each computational node on both
Hopper and Edison, with 32 and 64 gigabyte (GB) of memory per node,
respectively.


\subsection{Static calculations}\label{subsec:Force accuracy at GS}

We first examine the accuracy of the Hellmann-Feynman force for
quasi-1D and 3D disordered Si systems.

\begin{figure}
\centering
\includegraphics[width=0.45\textwidth]{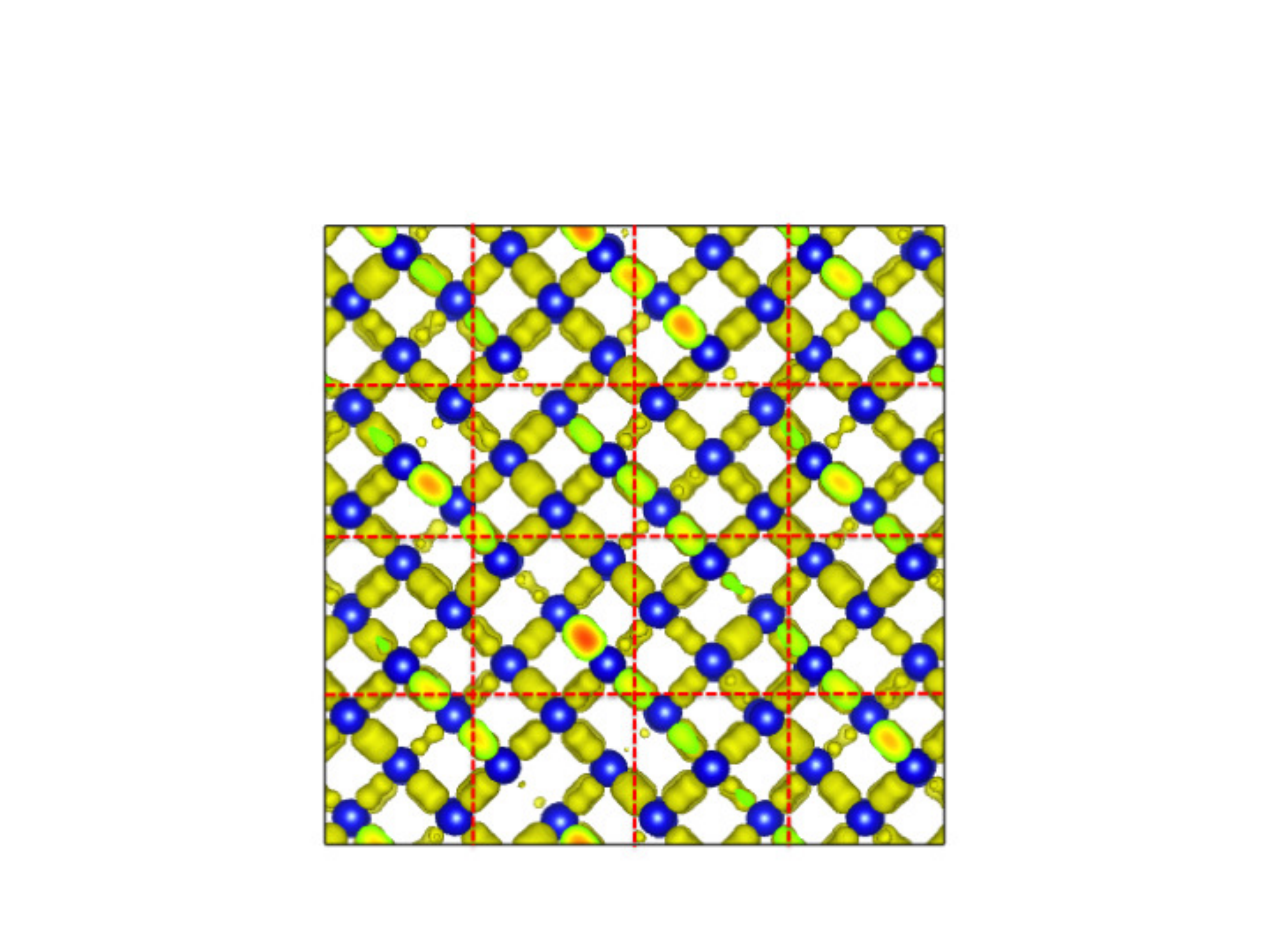}
\caption{(Color online) 3D-bulk disordered Si system with 216 atoms
is partitioned (red dashed lines) into 4 elements along
each direction, 64 elements in total, viewed along the $z$ direction.
Blue balls represent Si atoms, yellow surfaces represent the charge
density at a given isosurface, and green/orange surfaces represent
cross sections of charge density at boundaries. The disorder of the
structure is manifested in the differences between bonds.}
\label{fig:Si333}
\end{figure}

For the quasi-1D Si system, we replicate the 8-atom unit cell with
diamond structure along the $z$ direction 4 times, forming a
$1\times1\times4$ structure with 32 atoms in the supercell. The
$1\times1\times4$ Si system is partitioned into 6 elements along the
$z$ direction resulting in about 5 atoms per element. For the 3D
bulk Si system, we replicate the unit cell along $x$, $y$, and $z$
directions 3 times, to form a $3\times3\times3$ structure with 216
atoms in the supercell. The $3\times3\times3$ structure is
partitioned into 4 elements along each direction, giving 64 elements
in total. Each element contains around 3 atoms. The partition of the
$3\times3\times3$ Si structure is shown in Fig.~\ref{fig:Si333}. The
lattice constant of the Si unit cell is 10.2 au. The positions of
atoms in both Si systems are slightly perturbed by applying a random
displacement uniformly distributed within [-0.2,0.2] au.




To assess accuracy and convergence, we define the force error to be
\begin{equation}
\max_I |\Delta F_{I}|,
\label{eq:maxferr}
\end{equation}
where $I$ is the atom index,
\[
\Delta F_{I} = F_{I}^{\text{HF}} - F_{I}^{\text{ABINIT}},
\]
with $F_{I}^{\text{HF}}$ the Hellmann-Feynman force
computed by DGDFT and $F_{I}^{\text{ABINIT}}$ the
fully converged Hellmann-Feynman force computed by ABINIT.
The magnitude of the force error, denoted
by $|\Delta F_{I}|$, is then given by
\[
|\Delta F_{I}|=\sqrt{\Delta F_{I,x}^2+\Delta F_{I,y}^2+\Delta F_{I,z}^2},
\]
where $\Delta F_{I,x}$, $\Delta F_{I,y}$, and $\Delta F_{I,z}$ represent the
$x$, $y$, and $z$ components of $\Delta F_{I}$, respectively.

Excluding roundoff error, the force error defined by \eqref{eq:maxferr}
can be attributed to two factors:
1) numerical quadrature error that results from an insufficient number of grid
points required to perform numerical integration of several quantities such as
ALBs, density, and potential; and 2) an insufficient number of ALBs required to
accurately represent the Kohn-Sham wavefunctions. We shall refer the latter as
basis set error.  Note that the basis set error contributes
directly to the Pulay force through the incompleteness of the basis.

In order to examine the Pulay force and compare it with the error \eqref{eq:maxferr},
we first reduce the numerical quadrature error
by increasing the kinetic energy cutoff $E_{\text{cut}}$.
Note that increasing $E_{\text{cut}}$ also increases
$E_{\text{cut}}^{\text{den}}$ and $E_{\text{cut}}^{\text{LGL}}$
proportionally. The parameter $E_{\text{cut}}$ is eventually
constrained by the amount of computer memory available to store the ALBs
and their derivatives,
to a sufficiently large value so that numerical quadrature errors become negligible.
More specifically, we start from a large number of ALBs, and increase
the $E_{\text{cut}}$ defined in DGDFT gradually until the difference
between the total energy obtained from DGDFT and that obtained from a fully
converged ABINIT calculation is negligible.
Once the desired $E_{\text{cut}}$ is determined, we use this energy cutoff
in subsequent calculations to examine how the force error
changes with respect to the number of basis functions per element.

\subsubsection{Energy cutoff}\label{subsec:energy}

When DGDFT is used to compute the ground state energy and density,
we set the Fermi-Dirac smearing to
0.01 Ha, and DG penalty parameter to 20.
We terminate the SCF iteration when
\[
\| \rho_{\mathrm{out}} - \rho_{\mathrm{in}} \| / \| \rho_{\mathrm{in}} \| \leq 10^{-8},
\]
where $\rho_{\mathrm{in}}$ and $\rho_{\mathrm{out}}$ are the input and output
electron density in the SCF iteration, respectively.

To achieve convergence of the total energy in ABINIT, we increase
$E_{\text{cut}}$ until the change of the computed total energy
is below  $10^{-8}$ Ha per atom. We find that the smallest
$E_{\text{cut}}$ that achieves this level of convergence is 100 Ha.
The total energy at this $E_{\text{cut}}$ is then taken as the reference
for all subsequent error computations.

\begin{figure}
\centering
\includegraphics[width=0.6\textwidth]{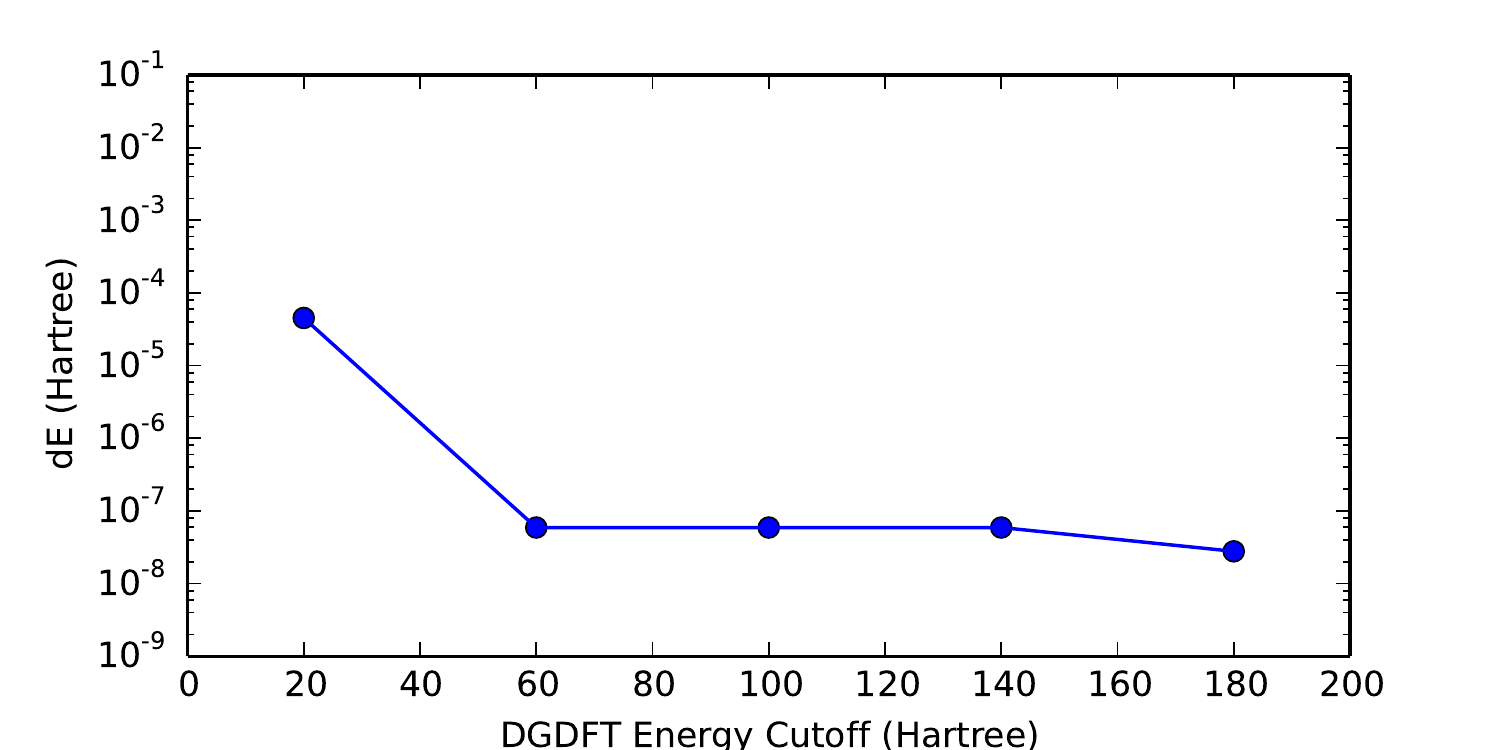}
\caption{(Color online) Difference between DGDFT and
 converged ABINIT total energy per atom with respect to
energy cutoff using a large number of ALBs per atom (45)
for the quasi-1D Si system.
}
\label{fig:si1d}
\end{figure}

In Fig.~\ref{fig:si1d}, we plot the difference between the total energy
computed by DGDFT with different DGDFT $E_{\text{cut}}$ values
and the converged ABINIT total energy for the quasi-1D Si system.
In these calculations, we use a large number of basis functions
($45$ ALBs per atom).
When the DGDFT $E_{\text{cut}}$ is increased to 60 Ha, the difference
between the DGDFT total energy  and the converged ABINIT total energy
is below $10^{-7}$ Ha per atom.
Figure~\ref{fig:si1d} indicates that
we may set $E_{\text{cut}}$ to 60 Ha in DGDFT
to make quadrature errors negligible for subsequent calculations.

\begin{table}
  \centering
  \begin{tabular}{cccc}
    \hline
    \hline
    System & \#ALB/atom & $E_{\text{cut}}$ (au)& $\Delta\mc{E}_{\text{tot}}$ per atom
    (au)\\
    \hline
    Quasi-1D disordered Si & $45$ & $60$ & $3\times 10^{-8}$\\
    3D disordered Si       & $47.4$ & $40$ & $3\times 10^{-5}$\\
    \hline
  \end{tabular}
  \caption{Accuracy of DGDFT total energy for test systems using a large
  number of ALBs and large kinetic energy cutoff.}
  \label{tab:accenergy}
\end{table}

We use the same procedure to identify the kinetic energy cutoffs
that make numerical quadrature error negligibly small for the 3D
disordered Si system. The number of ALBs per atom, determined
$E_{\text{cut}}$ values, and corresponding total energy errors are
reported in Table~\ref{tab:accenergy}.

\subsubsection{Pulay force}\label{subsec:force}

We use the $E_{\text{cut}}$ values determined from total energy convergence
tests and listed in Table~\ref{tab:accenergy} for the subsequent
investigation of the force computed by DGDFT with respect to the number
of adaptive local basis functions.
As opposed to the force error
$\Delta F$ that measures the difference between the force obtained from
DGDFT and the reference value obtained from ABINIT, the Pulay force characterizes the effect of
the atomic-position dependence of an incomplete basis set.

In order to compute the Pulay force $F_{I}^{\text{Pulay}}$ defined by
Eq.~\eqref{eqn:Pulay},
we use a second order finite difference method with
grid spacing $0.05$ au to compute the required derivatives to obtain
the force $F_{I}$ (Eq.~\eqref{eqn:forceDef}), and
subtract the Hellmann-Feynman force $F_{I}^{\text{HF}}$ (Eq.~\eqref{eqn:hfforce}). For all
examples, we report the Pulay force $F_{I}^{\text{Pulay}}$ for the atom
$I$ with the largest force error.

\begin{table}
 \caption{DGDFT force error ($\Delta F$) for quasi-1D Si system and Pulay force
($F^\text{Pulay}$) with respect to the number of adaptive local
basis functions per atom. \#ALB/atom indicates the number of
adaptive local basis functions per atom. Forces are in units of
Ha/Bohr.} \centering \scalebox{0.7}{
\begin{tabular}{cccccccc}
\hline\hline
\#ALB/atom & $\Delta F_x$ & $\Delta F_y$ & $\Delta F_z$
& $F^{\text{Pulay}}_x$ & $F^{\text{Pulay}}_y$ & $F^{\text{Pulay}}_z$
\\ [0.5ex]
\hline
    $9.0$ & $1.26\times10^{-4}$ & $7.38\times10^{-5}$ & $-6.55\times10^{-4}$ & $2.45\times10^{-4}$ & $-6.13\times10^{-5}$ & $-6.61\times10^{-4}$ \\
    $15.8$ & $-1.18\times10^{-5}$ & $3.00\times10^{-6}$ & $8.68\times10^{-5}$ & $-1.29\times10^{-5}$ & $-1.21\times10^{-5}$ & $5.08\times10^{-5}$ \\
    $25.5$ & $2.92\times10^{-6}$ & $-4.27\times10^{-7}$ & $-3.60\times10^{-6}$ & $1.82\times10^{-6}$ & $ 4.46\times10^{-6}$ & $4.00\times10^{-7}$ \\
\hline
\end{tabular}
}
\label{table:pf1}
\end{table}

The computed DGDFT force error~\eqref{eq:maxferr}
and Pulay force~\eqref{eqn:Pulay} for the quasi-1D Si system are shown in
Table~\ref{table:pf1} for several choices of \#ALB/atom.
We find that the force error of DGDFT decreases rapidly as the number of ALBs
per atom increases. The force error is
on the order of $10^{-5}$ au or below when approximately
16 ALBs per atom are used.
If we increase the number of ALBs further,
the force error can decrease to $10^{-6}$ au or smaller.
We also see from Table~\ref{table:pf1} that the computed Pulay forces
are on the same order as the total force error.
This observation suggests that, after reducing the numerical
quadrature error, most of the force error can be accounted for by the
Pulay force, which decreases rapidly with the number
of adaptive local basis functions used.
Therefore, the error of forces in DGDFT can be readily reduced to
the accuracy typically required in molecular dynamics and geometry
optimization applications ($\sim10^{-3}$ au) with $\sim$10 basis
functions/atom.


\begin{table}
\caption{DGDFT force error for 3D bulk Si system ($\Delta F$) and
Pulay force ($F^\text{Pulay}$) with respect to the number of
adaptive local basis functions per atom. \#ALB/atom indicates the
number of adaptive local basis functions per atom. Force errors are
in units of Ha/Bohr.} \centering \scalebox{0.7}{
\begin{tabular}{cccccccc}
\hline\hline
\#ALB/atom & $\Delta F_x$ & $\Delta F_y$ & $\Delta F_z$
& $F^{\text{Pulay}}_x$ & $F^{\text{Pulay}}_y$ & $F^{\text{Pulay}}_z$
\\ [0.5ex]
\hline
    $35.6$ & $3.72\times10^{-3}$ & $5.06\times10^{-3}$ & $6.25\times10^{-4}$ & $3.79\times10^{-3}$ & $7.14\times10^{-3}$ & $5.53\times10^{-3}$ \\
    $47.4$ & $1.66\times10^{-4} $ & $1.82\times10^{-4}$ & $6.15\times10^{-5}$ & $1.76\times10^{-4}$ & $1.73\times10^{-4}$ & $2.46\times10^{-5}$ \\
    $71.1$ & $2.65\times10^{-5} $ & $3.09\times10^{-5}$ & $1.03\times10^{-5}$ & $7.58\times10^{-5}$ & $6.28\times10^{-5}$ & $2.34\times10^{-5}$ \\
\hline
\end{tabular}
}
\label{table:pf3}
\end{table}

The same trend is observed for 3D bulk Si systems, as shown in
Table~\ref{table:pf3}. For the 3D bulk Si system, using around 47
ALBs per atom allows us to reduce the force error to $10^{-4}$
Ha/Bohr or below. The latter system requires a few times more basis
functions per atom due to the 3D element partition, consistent with
previous findings~\cite{LinLuYingE2012} for total energies.


\subsubsection{Vibrational frequencies}\label{subsec:vib}

Vibrational frequency is an important observable for characterizing
material properties at finite temperature.
For a system at equilibrium atomic configuration, the dynamical matrix is defined as
\[
D_{I;J} = \frac{1}{\sqrt{M_{I}M_{J}}}\frac{\partial^2
\mc{E}_{\text{tot}}}{\partial R_{I} \partial R_{J}},
\]
which is a square matrix of dimension $3 N_{A}$. The square root of its
eigenvalues gives the vibrational frequencies.
The computation of vibrational frequencies requires highly accurate
force calculations since the second order derivatives of the energy are
needed.

We compute the vibrational frequencies of the quasi-1D Si
system using the frozen phonon method.
The second derivatives of the total energy
$\frac{\partial^2 \mc{E}_{\text{tot}}}{\partial R_{I} \partial R_{J}}$
are computed by applying a central difference formula to the
Hellmann-Feynman forces computed by DGDFT. We note that the purpose
of the present calculation is to demonstrate the applicability of ALBs for
the computation of such second derivative information, rather than
to produce a physically meaningful vibrational spectrum, which would require
a considerably larger supercell and/or $k$-point sampling to obtain.

\begin{figure}
\centering
\includegraphics[width=0.6\textwidth]{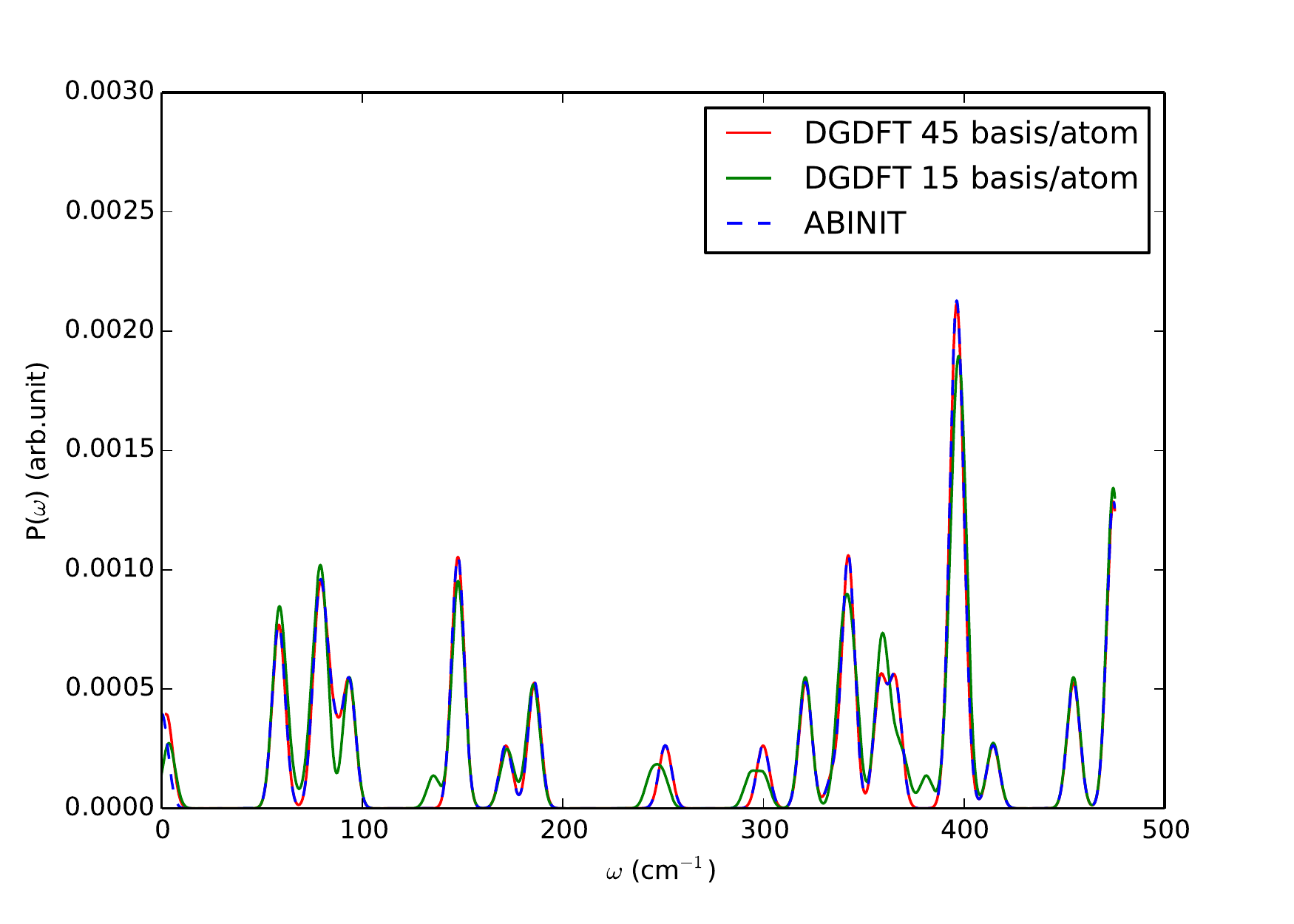}
\caption{(Color online) Vibrational frequency for quasi-1D Si
obtained from DGDFT using 45 basis functions per atom (red solid line) and
15 basis functions per atom (green solid line), and ABINIT
(blue dashed line).}
\label{fig:vf}
\end{figure}

To compute vibrational frequencies, we set the kinetic energy cutoff to 60 Ha
in both DGDFT and ABINIT, and use 15 and 45 ALBs per atom in
DGDFT. The displacement is set to $0.0102$ Bohr in each direction in
the finite difference method. As shown in Fig.~\ref{fig:vf},
we find that the vibrational frequencies obtained from DGDFT
converge to those from ABINIT as the number of ALBs is increased,
with excellent agreement at 45 ALBs/atom,
as expected given the agreement of DGDFT and ABINIT forces
shown in Section~\ref{subsec:force}.


\subsection{Molecular dynamics}\label{subsec:md}

We have shown for a range of static configurations that the DGDFT
force error can be as small as $10^{-4}$ Ha/Bohr with a moderate
number of adaptive local basis functions per atom. In this section,
we demonstrate that this level of accuracy, as for standard
planewave methods, yields converged \textit{ab initio} molecular
dynamics simulations as well. We consider two systems: one is a
molecular and the other a condensed matter system. The first system
consists of four H$_2$ molecules evenly spaced in a box of dimension
$10\times 10\times 40$ au. The other system is a liquid Al-Si alloy.
When performing molecular dynamics simulations in DGDFT, we
partition the four H$_2$ molecules into 4 elements of equal size
along the $z$ direction, and partition the Al-Si alloy into
$4\times4\times4$ elements of equal size. These two systems
constitute a quasi-1D structure and 3D structure, respectively.

\subsubsection{H$_2$}

The four H$_2$ molecules are simulated in the
constant-temperature-constant-volume (NVT) ensemble at 300 K, using
a Nose-Hoover thermostat~\cite{Nose1984,Hoover1985}. The integration
time step is 25 au ($\sim$0.6 fs), and the
thermostat masses $Q_s$ are set to 20 au for both DGDFT and ABINIT. In
DGDFT, we set the kinetic energy cutoff
to 160 Ha. 
With 10 adaptive local basis functions per atom, the force computed by
DGDFT is accurate to $10^{-6}$ au in the initial configuration.
We perform 2000 MD simulation steps in both ABINIT and DGDFT, 
which corresponds to 1.2 ps of simulation time.


\begin{figure}
\centering
\includegraphics[width=0.6\textwidth]{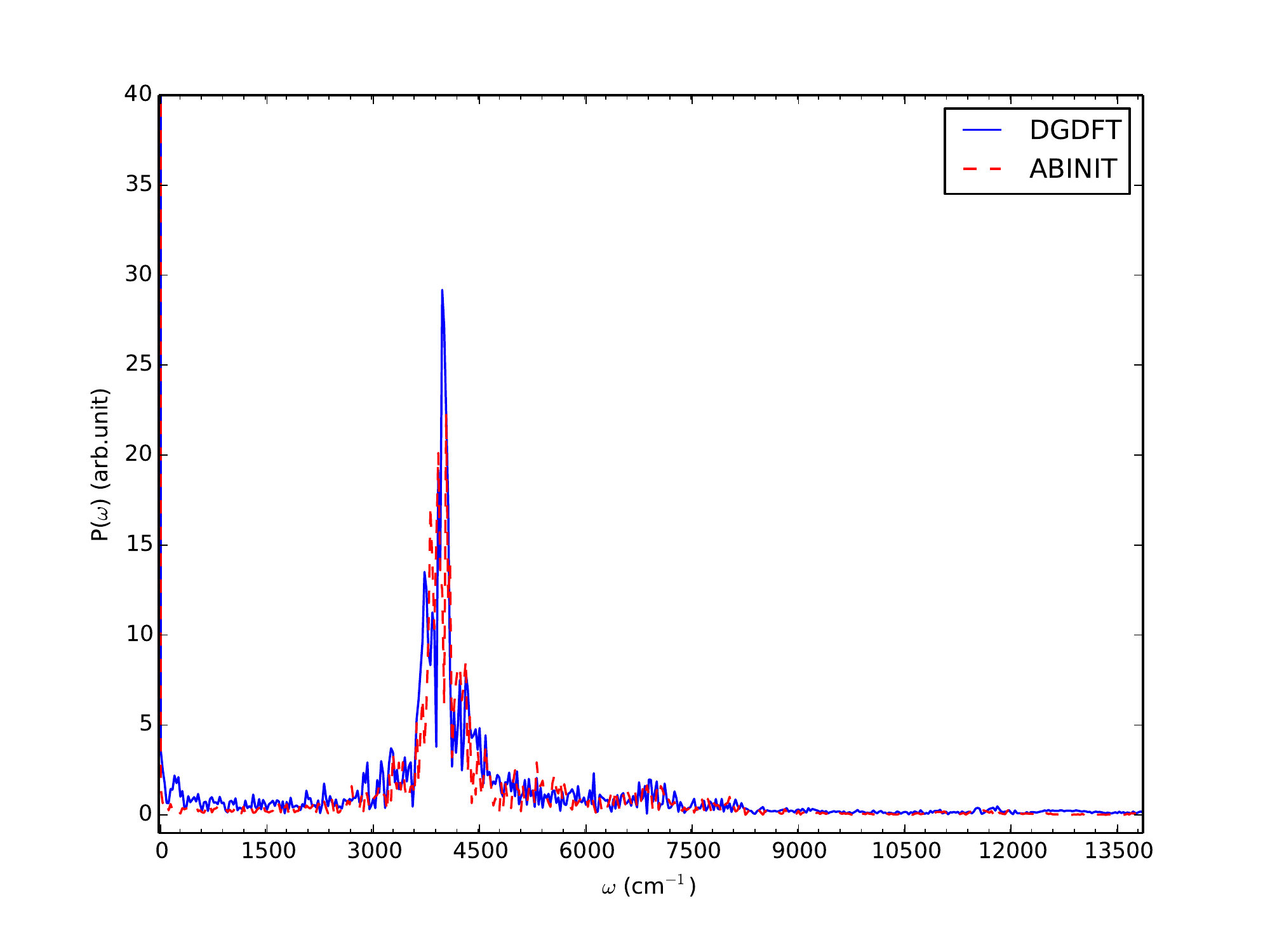}
\caption{(Color online) Vibrational frequency of H$_2$ molecule
obtained from DGDFT (blue solid line) and ABINIT (red dashed line)
at 300 K.} \label{fig:H2frequency}
\end{figure}


The vibrational spectrum is given by the Fourier transform of the
bond length of one H$_{2}$ molecule along the MD trajectory.
Fig.~\ref{fig:H2frequency} (blue solid line) shows the vibrational
spectrum calculated using DGDFT for the first H$_{2}$ molecule. The
general profile of the calculated spectrum agrees well with that
obtained from ABINIT (red dashed line). The characteristic
vibrational frequencies estimated by DGDFT and ABINIT differ by only
a few cm$^{-1}$. Although we obtain good quantitative agreement from
the 1.2 ps trajectories, longer trajectories may be required to
eliminate the multiple peaks around the characteristic frequency.

\subsubsection{Liquid Al-Si alloy}

Simulations of liquid Al-Si alloy Al$_{0.88}$Si$_{0.12}$ were performed
in a 200-atom cell with lattice constant 15.68 {\AA} in the canonical ensemble
at target temperature 973K, as in Ref.~\cite{KhooChanKimEtAl2011}.
In the initial configuration, Al and Si atoms are randomly
placed in the cubic cell.
The system is then relaxed at 4000 K,
well above the target temperature,
so that any biases in the initial configuration are removed. We then
cool down the system from 4000 K to 973 K gradually, at the rate of
0.471 K/fs, and continue the simulation in the canonical ensemble at
973 K for more than 2.0 ps to equilibrate.
The simulation continues under these conditions for another 7.0 ps,
and data is collected during this period to compute the properties
of the Al-Si alloy.
We use a Nose-Hoover chain thermostat. In DGDFT, we set the
integration time step for the equation of motion to 100 au ($\sim$
2.42 fs), the thermostat mass to 10000 au, the energy cutoff to 10
Ha, and employ a basis of 51.2 adaptive local basis functions per
atom on average. The Nose-Hoover thermostat has a conserved energy,
which can be used to check the consistency of energies and forces.
We find a drift of the conserved energy of 2.6 meV/ps/atom,
consistent with the high accuracy of forces.



%

We next compare the statistical properties of the liquid Al-Si alloy
obtained from DGDFT with those obtained from
SIESTA~\cite{SolerArtachoGaleEtAl2002}, and results presented in
Ref.~\cite{KhooChanKimEtAl2011}, obtained from the PARSEC electronic
structure code. In particular, we examine the pair correlation
function, a statistical quantity widely used to characterize liquid
structure and coordination.
For alloy systems, the total pair correlation is given by:
\begin{equation}\label{eqn:Faber-Ziman}
  g^{\text{tot}}(r)=\frac{1}{b^2}({x_i}^2{b_i}^2g_{ii}(r)+2x_ix_jb_ib_jg_{ij}(r)+{x_j}^2{b_j}^2g_{jj}(r))
\end{equation}
according to the Faber-Ziman formalism~\cite{FaberZiman1965}. Here,
the indices $i$ and $j$ indicate different types of atoms, $x_i$,
$x_j$ are the corresponding molar fractions, $b_i$, $b_j$ are
neutron scattering lengths or x-ray form factors, $g_{ij}(r)$ are
the partial pair correlation functions, and
$b={x_i}{b_i}+{x_j}{b_j}$. For Al and Si, we use the neutron
scattering length ratio $b_{Al}/b_{Si} = 0.8318$.

\begin{figure}
\centering
\includegraphics[width=0.45\textwidth]{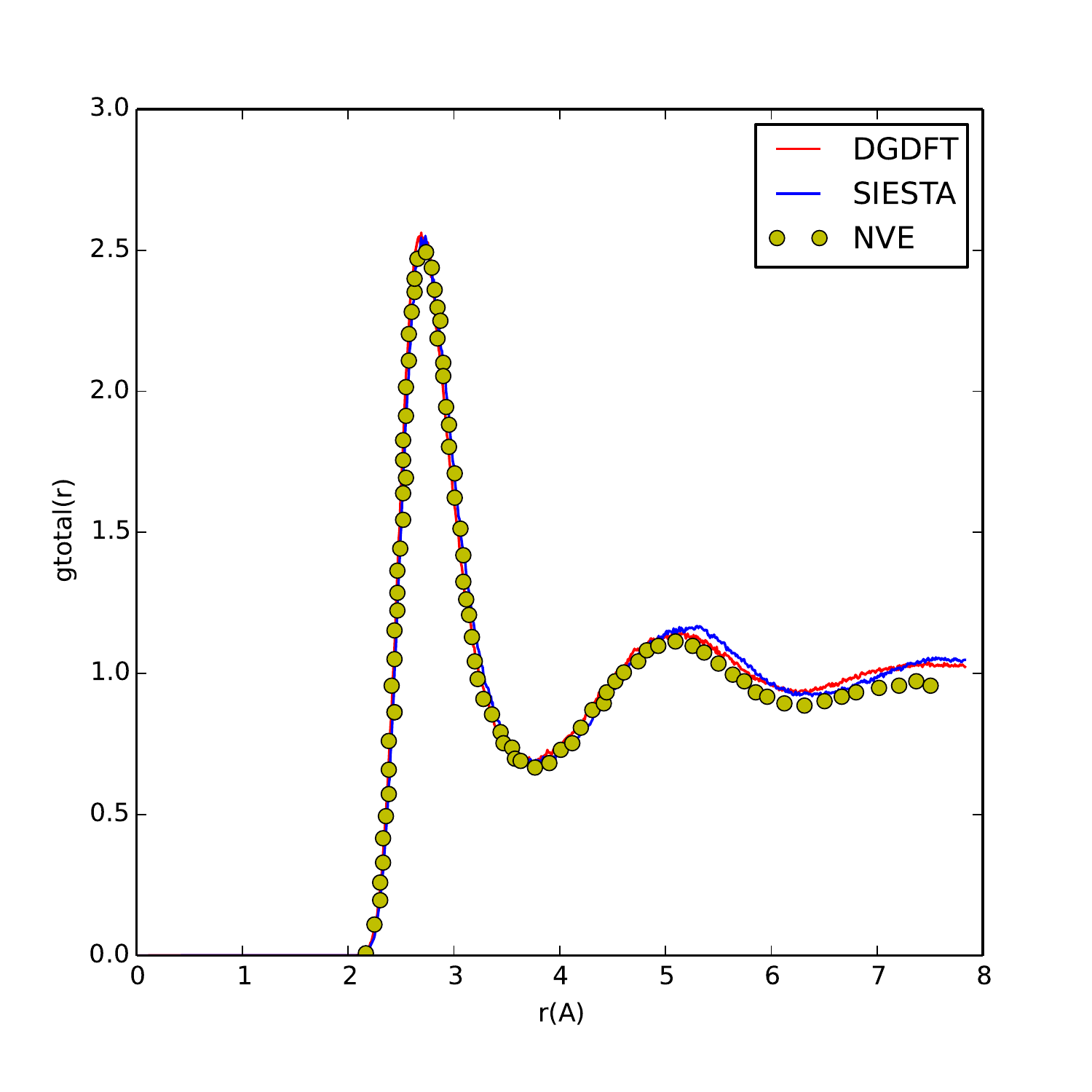}
\caption{(Color online) Comparison of total pair correlation function of liquid $Al_{0.88}Si_{0.12}$
calculated at 973 K from 200-atom simulations: DGDFT (red solid line), SIESTA (blue solid line),
and previous work~\cite{KhooChanKimEtAl2011} (yellow dots).}
\label{fig:paircorrelation}
\end{figure}

In Fig.~\ref{fig:paircorrelation}, we show the total pair
correlation functions $g(r)$ computed using DGDFT and SIESTA, and
results from Ref.~\cite{KhooChanKimEtAl2011} (labeled NVE). The
curve produced by DGDFT matches well with the other two results,
especially near the first peak. The liquid structure and
coordination are thus well described by all three codes. Slight
differences among the three are likely due to differences in
pseudopotentials, basis sets, and ensembles used.


\begin{figure}
\centering
\includegraphics[width=0.95\textwidth]{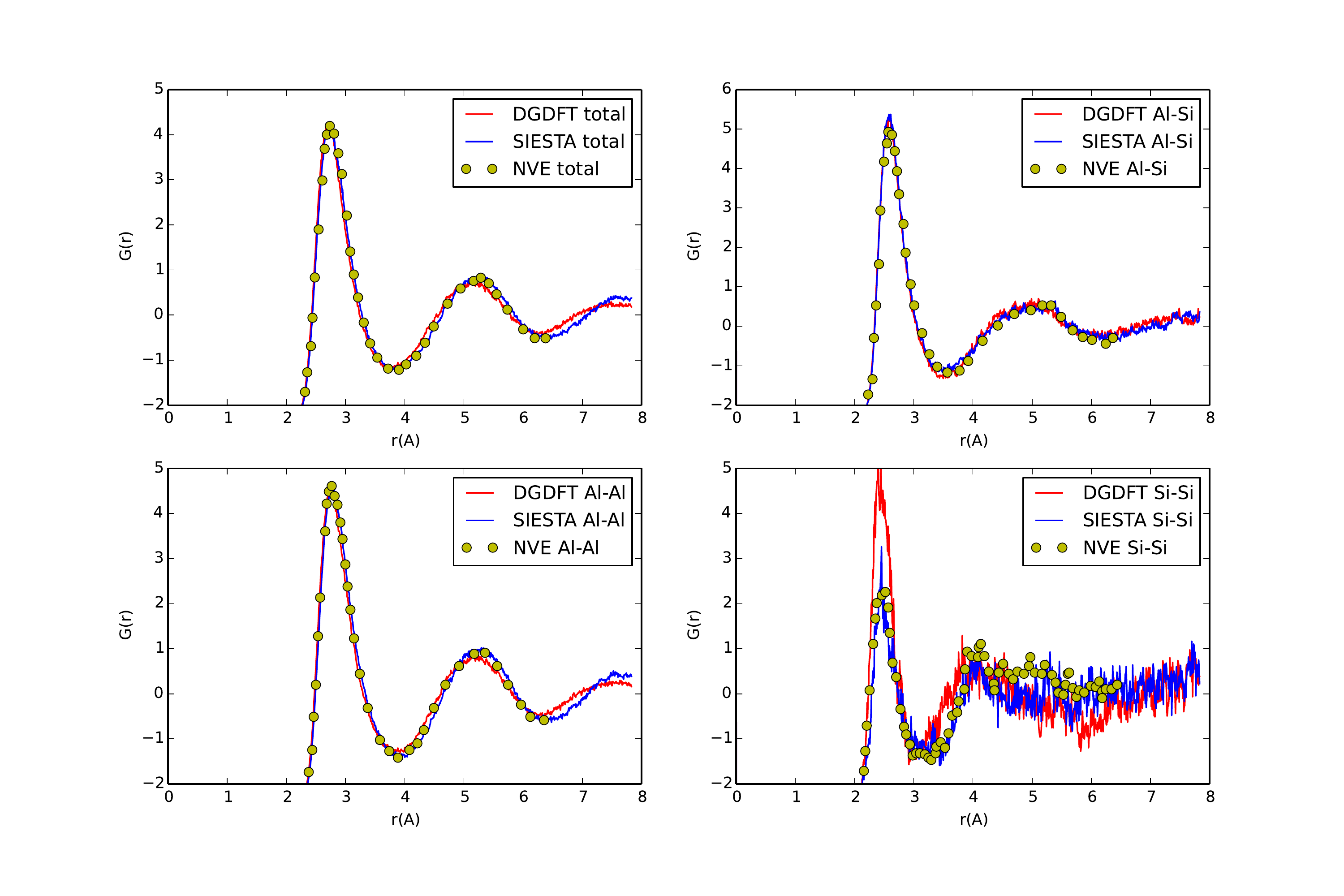}
\caption{(Color online) $G(r)$ derived from partial and total pair correlation functions
for liquid $Al_{0.88}Si_{0.12}$ calculated at 973K from 200-atom simulations using DGDFT (red
solid line) and SIESTA (blue solid line), and from 500-atom simulations in previous
work~\cite{KhooChanKimEtAl2011} (yellow dots).}
\label{fig:AlSiG}
\end{figure}

In Ref.~\cite{KhooChanKimEtAl2011}, the functions $G(r)=r(g(r)-1)$
and $G_{ij}(r)=r(g_{ij}(r)-1)$ are considered.  We plot the
corresponding quantities in Fig.~\ref{fig:AlSiG}. The $G(r)$,
$G_{Al-Al}(r)$, and $G_{Al-Si}(r)$ obtained by all three codes agree
well. The agreement of the curves, especially up to the first
minimum, indicates that DGDFT describes the Al-Si and Al-Al bond
lengths and coordinations well. In addition to the differences in
basis sets used in three methods, the number of atoms used in the
DGDFT and SIESTA simulations is different from that used in the
PARSEC simulations. The result in Ref.~\cite{KhooChanKimEtAl2011} in
Fig.~\ref{fig:AlSiG} used a 500-atom Al-Si system, while 200 atoms
were used in the DGDFT and SIESTA simulations. The difference in
number of atoms can also contribute to the slight differences in
computed pair correlation functions at larger $r$. Unlike the other
partial pair correlation functions, the differences in
$G_{Si-Si}(r)$ among the three methods is more noticeable. As noted
in Ref.~\cite{KhooChanKimEtAl2011}, the uncertainties are primarily
a result of insufficient statistics for the Si, which occurs in much
smaller number than Al in the present alloy.

\section{Conclusion}\label{sec:discussion}


In this work, we detail the calculation of atomic forces in
Kohn-Sham density functional theory using an adaptive local basis.
We demonstrate that, under mild assumptions, the computation of
atomic forces can scale nearly linearly with the number of atoms in
the system using the adaptive local basis set. The method is
implemented in the recently developed DGDFT software package which
achieves high efficiency on massively parallel computers. We
quantify the accuracy of the Hellmann-Feynman forces for a range of
physical systems, benchmarked against converged planewave
calculations, and find that the adaptive local basis set is
efficient for both force and energy calculations, requiring at most
a few tens of basis functions per atom to attain accuracies required
in practice. Since the adaptive local basis set has implicit
dependence on atomic positions, Pulay forces are in general nonzero.
However, by virtue of the adaptive local construction, we find that
the Pulay force is small and systematically decreasing with
increasing basis completeness; so that the Hellmann-Feynman force is
sufficient for basis sizes of a few tens of basis functions per
atom. We verified the accuracy of computed Hellmann-Feynman forces
in static calculations of quasi-1D and 3D Si systems, finding
convergence of forces to $10^{-4}$ Ha/Bohr with at most a few tens
of basis functions per atom in all cases. We further verified the
accuracy of the computed forces in frozen phonon calculations of the
vibrational spectrum of quasi-1D Si, and molecular dynamics
simulations of H$_2$ molecules and liquid Al-Si alloy, finding in
all cases excellent agreement with independent calculations and
benchmark results.

Whereas the Hellmann-Feynman force is sufficient for adaptive local
bases of a few tens of basis functions per atom or more, if still
smaller bases are desired, as for example in density-matrix based
$\Or(N)$ calculations, then Pulay forces will become significant. In
such case, optimized local basis functions
(OLB)~\cite{LinLuYingE2012a} may be a better candidate than adaptive
local basis functions (ALB) since the Pulay force could be reduced
systematically and substantially. However, the optimized local basis
functions introduce other numerical difficulties in implementation,
particularly for 3D systems. We plan to investigate this in the
future.



\section*{Acknowledgments}

This work was supported by the Scientific Discovery through Advanced
Computing (SciDAC) program funded by U.S. Department of Energy,
Office of Science, Advanced Scientific Computing Research and Basic
Energy Sciences (G. Z, L. L., W. H., C. Y. and J. P.), and the
Center for Applied Mathematics for Energy Research Applications
(CAMERA), which is a partnership between Basic Energy Sciences (BES)
and Advanced Scientific Computing Research (ASRC) at the U.S
Department of Energy (L. L. and C. Y.). This work was performed, in
part, under the auspices of the U.S. Department of Energy by
Lawrence Livermore National Laboratory under Contract
DE-AC52-07NA27344. We thank the National Energy Research Scientific
Computing (NERSC) center for providing the computational resources.


\appendix

\section{Appendix: Treatment of electrostatic interactions in extended systems}

In this appendix, we discuss how extended periodic systems can be
treated, and provide corresponding expressions for the total energy
and atomic forces. The most widely used method for treating periodic
systems relies on the use of Ewald summation for long-range
interactions, which appear both in the local part of the
pseudopotential and in the ion-ion interaction.  An alternative
approach is to use the pseudocharge formulation as given in
Eq.~\eqref{eqn:pseudocharge} to replace long-range ionic potentials
by corresponding short-range charge densities, eliminating the need
for Ewald summation
altogether~\cite{SolerArtachoGaleEtAl2002,PaskSterne2005}. The
advantage of such a formulation is that the electrons and ions are
treated on the same footing, in a single Poisson solution,
permitting both electron-ion and ion-ion interactions to be
evaluated at once in $\Or(N)$ operations.

The total ionic pseudocharge in a unit cell is
the sum of pseudocharges from all atoms:
\begin{equation}
  \rho_{\text{loc}}(x) = \sum_{I=1}^{N_A} \rho_{\text{loc},I}(x-R_{I}),
  \label{eqn:totalrholoc}
\end{equation}
where $N_A$ includes all atoms with nonvanishing
pseudocharge $\rho_{\text{loc},I}$ in the unit cell.
The Kohn-Sham energy functional can then be written as
\begin{equation}
  \begin{split}
     &\mc{E}_{\KS}(\{\psi_i\};\{R_{I}\}) \\
    =& \frac{1}{2} \sum_{i=1}^{N} \int \abs{\nabla
    \psi_i(x)}^2 \ud x
    + \sum_{i=1}^{N} \int \psi_{i}(x) V_{\text{nl}}(x,y;\{R_{I}\})
    \psi_i(y) \ud x \ud y \\
    &+\frac{1}{2} \int (\rho(x)-\rho_{\text{loc}}(x)) V_C(x) \ud x
    + E_{\xc}[\rho]
    - E_{s}.
  \end{split}
  \label{eqn:KSperiodic}
\end{equation}
Since
$\int \rho(x)-\rho_{\text{loc}}(x)\ud x = 0$
in the unit cell, the total Coulomb potential $V_C$ due to ionic and electronic charge densities
can be obtained by solving the Poisson equation
\begin{equation}
  -\Delta V_C(x) = 4\pi (\rho(x) - \rho_{\text{loc}}(x))
  \label{eqn:VHtot}
\end{equation}
subject to periodic boundary conditions on the unit cell.
$E_s$ is the self-interaction energy of pseudocharges $\rho_{\text{loc},I}$
in the unit cell, a constant independent of atomic positions~\cite{PaskSterne2005}.
Having computed the total Coulomb potential $V_C$,
the Hellmann-Feynman force is then given by
\begin{equation}
    F_{I}^{\text{HF}} = \int \rho_{\text{loc},I}(x-R_{I}) \nabla_{x}
    V_C(x) \ud y
     + 2 \sum_{\ell=1}^{L_{I}}\Tr[V_{\text{nl},I,\ell} P].
  \label{eqn:hfforceperiodic}
\end{equation}
Compared to Eq.~\eqref{eqn:hfforce}, the force due to the ion-ion
interaction is taken into account by the total Coulomb potential.


\end{document}